\documentclass[12pt,a4paper]{article}

\usepackage{amsmath}
\usepackage{amsfonts}
\usepackage{amsthm}

\def\R{\mathbb R}
\def\dom{\mathrm {dom}\,}
\def\LL{\mathrm L}
\def\CC{\mathrm C}
\def\Id{\mathbf 1}

\newenvironment{dm}[1][Proof: ]{\textbf{#1}}{}
\newcommand{\cq}{\hfill \rule {2.5mm} {3mm}\vspace{0.5cm}}

\theoremstyle{definition}
\newtheorem{defi}{Definition}[section]

\newtheorem{lem}[defi]{Lemma}
\newtheorem{teo}[defi]{Theorem}

\begin{document}

\title{On the spectrum and weakly effective operator for Dirichlet Laplacian in thin deformed tubes}
\author{C\'esar R. de Oliveira\thanks{Corresponding author. Email: oliveira@pq.cnpq.br. Telephone: +55 16 8135 0039. Fax: +55 16 3351 8218.} {\small\; and\;} Alessandra A. Verri\\
\vspace{-0.6cm}
\small
\em Departamento de Matem\'{a}tica -- UFSCar, \small \it S\~{a}o Carlos, SP,
13560-970
Brazil\\ \\}
\date{\today}

\maketitle

\begin{abstract}  We study the Laplacian in deformed thin (bounded or unbounded)
tubes in~$\R^3$, i.e., tubular
regions along a curve $r(s)$ whose cross sections are multiplied by an appropriate deformation function~$h(s)> 0$. One the main requirements  on $h(s)$ is that it has a
single point of global maximum. We find the asymptotic behaviors of the
eigenvalues and weakly effective operators as the diameters of the tubes tend to zero. It is shown that
such behaviors are
not influenced by some geometric features of the tube, such as curvature,
torsion and twisting, and so a huge amount of different deformed tubes are
asymptotically described by the same weakly effective operator.
\end{abstract}

\

\noindent {\bf Keywords:} spectrum; thin tubes; Laplacian; dimensional reduction.

\

\noindent {\bf MSC codes:}  81Q15, 49R50, 35P20, 47B99

\section{Introduction}

The Laplacian in tubular domains has been studied in various situations
\cite{Bou, Cr1, DE, Kre}.
A common   tubular region~$\Omega$ is as follows:
let
$I \subseteq \mathbb R$ be an interval of $\mathbb R$,
$r:I  \to \mathbb R^3$ a curve in $\mathbb R^3$, parametrized by its arc length~$s$,
and
$k(s)$ and $\tau(s)$ denote its curvature and torsion at the point $s
\in I$, respectively.
Let $S$ be an open, bounded, simply connected and nonempty subset of
$\mathbb R^2$.
Move
the region~$S$
along~$r(s)$ and at each point~$s$ allow the region to rotate by an
angle~$\alpha(s)$  (see details in Section~\ref{gddh}).
A problem of interest is the description of the spectral properties of the Laplacian in such
tubes and weakly effective operators (see the definition just after Theorem~\ref{mainTeor})  when  the region~$\Omega$ is ``squeezed''  to the curve~$r(s)$, that is, one considers the
sequence of tubes~$\Omega_\varepsilon$
generated by the cross section~$\varepsilon S$ and analyze the limit~$\varepsilon \to 0$.

Let  $-\Delta_\varepsilon$ be the Dirichlet Laplacian  in 
$\Omega_\varepsilon$.
For bounded tubes, i.e., when  $I$ is a bounded interval  of $\mathbb R$,
the spectrum of  $-\Delta_\varepsilon$ is   purely discrete  because in
this case
its
resolvent is compact.
In \cite{Bou} it was analyzed the convergence of the eigenvalues
$\{\lambda_i^\varepsilon: i \in \mathbb N\}$
as  $\varepsilon \to 0$ and
shown that
$$\lambda_i^\varepsilon = \frac{\lambda_0}{\varepsilon^2} +
\mu_i^\varepsilon, \qquad \mu_i^\varepsilon \to \mu_i,$$
where $\lambda_0$ is the first, i.e., the lowest, eigenvalue of the Laplacian in the Sobolev
space
${\mathcal H}_0^1(S)$, and  $\mu_i$ are the eigenvalues of the
one-dimensional operator
\begin{equation}\label{bbb8}
w(s)\;\mapsto \; - w''(s) + \left[ C(S) (\tau(s) + \alpha'(s)) -
\frac{k(s)^2}{4} \right] w(s),
\end{equation} acting in~$\LL^2(I)$.
Here $C(S)$ is a nonnegative number depending only on the transverse
region~$S$~\cite{Bou}.
Note that this  effective operator explicitly depends on the geometric
shape of the reference curve~$r(s)$ (and so of the tube).

An interesting problem is to know if there exists a similar result about
convergence of
eigenvalues for unbounded tubes.
For such tubes, in~\cite{Kre}  it is shown that 
if $(\tau + \alpha')(s) = 0$ and  $k(s) \neq 0$, then the discrete
spectrum is nonempty, whereas
if  $(\tau + \alpha')(s) \neq 0$ and $k(s) = 0$, then the discrete spectrum is
empty.
In~\cite{Cr1}, by using
$\Gamma$-convergence in case of unbounded tubes, a strong resolvent convergence
was proven  and the same action \eqref{bbb8} for the respective effective
operator (now acting in $\LL^2(\mathbb R)$) was found as~$\varepsilon \to
0$.

The Dirichlet Laplacian in strips of $\mathbb R^2$ has been studied in
many works \cite{BF,Frie, Sol,Krej2}. For the  case of the constraints  of planar motion  to curves  there are results about  the limit operator in
\cite{DellAntTen,ACF}, and the effective potential is written in terms of the curvature.
The main
novelty, when we pass from planar domain  to tubes in~$\R^3$, as considered in \cite{Bou,Cr1,Kre, GJ}, is the additional presence of torsion and twisting (i.e., a nonzero $\tau(s) + \alpha'(s)$)  in the
effective potential, since the case of untwisted tubes has also been previously studied (see, for instance, \cite{DE,CDF,ClaBra,Ekk,FreitasKrej,KS}). 

In \cite{Frie,Sol} the authors consider  a family of deformed strips
\[
\left\{(s, y) \in \mathbb R^2: s \in J, \,0 < y < \varepsilon h(s) \right\},
\]
where  $J = [-a, b]$, $0< a, b \leq \infty$, and
$h(s) > 0$ is a continuous function satisfying:
\begin{itemize}
\item[(i)] $h(s)$ is a $\CC^1$ function in $J\setminus\{0\}$ and $\|h'/h\|_\infty<\infty$;
\item[(ii)] near the origin $h$ behaves  as
\begin{equation}\label{vvv1}
h(s) = M - s^2 + O( |s|^3), \qquad M > 0,
\end{equation}
and  $s=0$ is a single point of global maximum for~$h$;
\item[(iii)] in case $I=\R$ it is assumed that $  \limsup_{|s| \to \infty} h(s) < M$.
\end{itemize}
In what follows we assume that $h$ satisfies the above conditions.

It was shown  \cite{Frie,Sol} that, for  $\varepsilon$ small enough,  the discrete spectrum
of the Laplacian is
always nonempty and the eigenvalues
$\lambda_j(\varepsilon)$ have the following behavior
$$
\mu_j = \lim_{\varepsilon \to 0}\, \varepsilon \left(
\lambda_j(\varepsilon) -
\frac{\pi^2}{\varepsilon^2 M^2} \right),$$
where $\mu_j$ are the eigenvalues of the operator in
$\LL^2(\mathbb R)$ (it acts on a subspace of~$\LL^2(\mathbb R)$,
independently if the interval~$I$ is bounded or not)
given by
$$(T w)(s) = - w''(s) + 2 \frac{\pi^2 }{M^{3}} \; s^2 \; w(s),$$
so that we say that~$T$ is a {\em weakly effective operator} (WEO) in such situation.

In this work we show that these  results  hold in a more
general setting.
We consider a sequence of  tubes~$\Omega_\varepsilon$  in the
space~$\mathbb R^3$, as presented at the beginning of this
introduction, but we deform them by multiplying their cross sections by the above function~$h(s)$.
Here the tubes may be bounded or not.
Then we analyze  the asymptotic behavior of eigenvalues  and the weakly effective operators  in the limit~$\varepsilon\to0$.
The situation here differs from \cite{Frie,Sol}, since besides the
different dimensions (we consider regions in $3-$dimensional space), the reference curves defining  our tubes are allowed to
have nontrivial curvatures and torsion.
These tubes, which we shall call {\em deformed tubes},  will generically be
denoted by~$\Lambda_\varepsilon$ (see details in Section~\ref{gddh}).

Our main goal is to study how curvature and torsion of
the reference curve, together with the deforming function $h$, influence the WEO and eigenvalues as~$\varepsilon \to 0$. To this end, we introduce some notation right now.
Recall that~$\lambda_0$ is the lowest eigenvalue of the negative
Laplacian with Dirichlet conditions in the region~$S$, and let $u_0$ be the corresponding (positive) normalized eigenfunction, that is,
\begin{equation}\label{eqU0lambda0}
-\Delta u_0 = \lambda_0 u_0, \quad u_0 \in {\mathcal H}_0^1(S),
\quad \int_S u_0(y)^2 dy = 1.
\end{equation} Furthermore, denote by~${\mathcal L}$ the subspace of~$\LL^2(I\times S)$ generated by functions
$w(s) u_0(y)$ with $w \in \LL^2(I)$.

We study three distinct cases. First, the tubes are bounded since the
interval~$I$ is of the form
$I=[-a,b]$ with $0< a, b < \infty$, and we consider the Dirichlet
condition at the
boundary
$\partial \Lambda_\varepsilon$.
In the second case, the tubes are bounded  but the Dirichlet condition at
the vertical part of the $\partial \Lambda_\varepsilon$, that is, 
$\{(-a) \times S \cup b \times S\}$,
is replaced by Neumann. In the third case we consider  $I=\mathbb R$ with
Dirichlet
condition at~$\partial \Lambda_\varepsilon$.

If the tubes are not deformed, according to the results of  \cite{Bou,Cr1},  the effective operator~\eqref{bbb8}
presents an additional potential
\[
C(S)\left(\tau+\alpha'\right)(s) - k^2(s)/4
\] derived from  geometric features
of the tube. Hence, here there is a kind of competition  between geometric
properties of the tube and the behavior at its single maximum of the
deformation function~$h$. Roughly speaking, it is expected that the behavior of $h$ at the single maximum will control the limit $\varepsilon\to0$, since the geometric effects gives a contribution of order zero, whereas the single maximum of the deformation function~$h$ gives a contribution of order $1/\varepsilon$. However, this requires a proof which turns out to be far from trivial, and so  for the three cases mentioned in the
previous paragraph, we prove the following result:

\

\begin{teo}\label{mainTeor}
Let $I$ denote either $\R$ or a bounded interval $[-a,b]$ as above; in case $I=\R$ assume that $\lim_{|s| \to \infty} k(s) =0$. If $l_j(\varepsilon)$ denote the eigenvalues of the Dirichlet
$-\Delta_\varepsilon$ in the deformed tube~$\Lambda_\varepsilon$, then,
the limits
\begin{equation}\label{eee3}
\mu_j = \lim_{\varepsilon \to 0}\, \varepsilon \left( l_j(\varepsilon) -
\frac{\lambda_0}{\varepsilon^2 M^2} \right)
\end{equation}
exist, where  $\mu_j= (2j+1)(2\lambda_0/M^3)^{1/2}$ are the eigenvalues of the self-adjoint
operator~$T$, acting in $\LL^2(\mathbb R)$, given by
\begin{equation}\label{eqweo}
(T u)(s) = - u''(s) +  2\frac{ \lambda_0 }{ M^{3}} \;  s^2 \; u(s).
\end{equation}
\end{teo}

\

Due to the conclusions of Theorem~\ref{mainTeor},  $T$ is a WEO for $-\Delta_\varepsilon$ as $\varepsilon\to0$. Note that $T$ has purely discrete spectrum since the potential 
\[
V(s)=2\frac{\lambda_0}{M^3}\, s^2\to\infty,\qquad |s|\to\infty;
\] in this case it is the harmonic oscillator potential (but see~\eqref{aaa6} below). Therefore, for deformed tubes as above, the weakly effective operators~$T$ do not
depend on some geometric features of the tube, although the curvature of the reference curve must vanish at infinity. The additional potential $V(s)$ is related to the behavior of~$h(s)$
near it maximum (at the origin). Hence, the eigenvalues of the Laplacian in quite different deformed tubes are described by the same WEO as~$\varepsilon\to0$ !

In Section~\ref{gddh} we present a detailed construction of the
deformed tubes~$\Lambda_\varepsilon$. Our study and technique are  focused on
analyzing the sequence of quadratic forms
\begin{equation}\label{seqfq}
F_\varepsilon (\psi) = \int_{\Lambda_\varepsilon} \left(|\nabla \psi|^2
- \frac{\lambda_0}{\varepsilon^2 M^2} |\psi|^2\right) dx,
\qquad \dom F_\varepsilon = {\mathcal H}_0^1(\Lambda_\varepsilon).
\end{equation}
In Section~\ref{ffqq} it will become clear why we subtract terms of the
form  $\lambda_0 / (\varepsilon^2 M^2)|\psi|^2$ from the quadratic forms; we think
this is in fact a natural choice. In Section~\ref{ffqq} we also perform a
change of variables so that the integration region and the corresponding
domains in~\eqref{seqfq} remain fixed.
In Section~\ref{redim}, we show that our analysis can be restricted to a
specific subspace;
we will see that this subspace can be identified  with the Sobolev space
${\mathcal H}_0^1(I)$, and  we call this fact a
{\em reduction of dimension.}
Finally, in  Sections~\ref{qqq3}, \ref{nnnn} and~\ref{iiii}, we discuss
details of the three cases previously
mentioned.

We remark that although we rely on \cite{Frie,Sol}, the generalization to our setting is not immediate and different techniques are added to those of the original works. Furthermore, as an alternative to~\eqref{vvv1}, all results can be easily adapted to cover more general 
deformation functions $h(s)$, as considered in \cite{Frie,Sol}, so that
near the unique global maximum at the origin they behave as
\begin{equation}\label{aaa6}
h(s) = \left\{
\begin{array}{ccc}
M - c_+ s^m + O(s^{m+1}),  & \hbox{if}  & s>0 \\
M - c_- |s|^m + O(|s|^{m+1}),  & \hbox{if}   & s<0
\end{array}\right.,
\end{equation}
for some positive numbers $M,m,c_\pm$.
For the sake of simplicity, in Equation~\eqref{vvv1} we have particularized to $m=2$ and
$c_+=c_-=1$.

An interesting problem would be if the  maximum of~$h$ would be reached at an interval of values of the parameter~$s$ instead of a single point (see~\cite{BF2} for results in this direction in case of bounded domains, as well as~\cite{BKRS,Grushin}); we are currently working on a related problem.

\section{Geometry of the tubes}\label{gddh}

Let $I=[-a, b]$, with either $0 < a,b< \infty$ or $a=b=\infty$, be an interval of  $\mathbb R$,
$r: I \subseteq \mathbb R  \rightarrow \mathbb R^3$ a simple  $\CC^2$ curve in
$\mathbb R^3$ parametrized by its arc length parameter~$s$ and, as in the previous section, $k(s)$ is its curvature.
The vectors
$$T(s) = r'(s), \qquad N(s) = \frac{1}{k(s)} T'(s), \qquad
B(s) = T(s) \times N(s),$$
denote, respectively, the tangent, normal and binormal vectors of the curve.
We assume that Frenet equations are satisfied, that is,
$$\left(\begin{array}{c}
T'\\
N'\\
B'
\end{array} \right)
=
\left(\begin{array}{ccc}
0 & k & 0 \\
-k & 0 & \tau \\
0 & -\tau & 0
\end{array}\right)
\left(\begin{array}{c}
T\\
N\\
B
\end{array} \right),
$$
where $\tau(s)$ is the torsion of the curve~$r(s)$.

Let  $S$ be an open, bounded, simply connected and nonempty subset of~$\mathbb R^2$.
The set
$$\Omega = \left\{ x \in \mathbb R^3: x = r(s) + y_1 N(s) + y_2 B(s),\,
s \in I, y=(y_1, y_2) \in S \right\}$$
is obtained by translating the region $S$ along the curve~$r$. At each point
$r(s)$ we allow a rotation of the region~$S$  by an angle $\alpha(s)$ with
respect to $\alpha(0)=0$, so that the new region is given by
$$\Omega^\alpha = \left\{ x \in \mathbb R^3: x = r(s) + y_1 N_\alpha(s) + y_2
B_\alpha(s),
\,s \in I, (y_1, y_2) \in S \right\},$$
where
\begin{eqnarray*}
N_\alpha(s) & : = & \cos \alpha(s) N(s) + \sin \alpha(s) B(s), \\
B_\alpha(s) & : = & - \sin \alpha(s) N(s) + \cos \alpha(s) B(s).
\end{eqnarray*}
Next, for each  $0<\varepsilon <1$, we ``squeeze'' the cross sections of the above region,
that is, we consider
$$\Omega_\varepsilon^\alpha = \left\{ x \in \mathbb R^3: x = r(s) + \varepsilon
y_1 N_\alpha(s) + \varepsilon y_2 B_\alpha(s),\,
s \in I, (y_1, y_2) \in S\right \}.$$
Note that $\Omega_\varepsilon^\alpha$  approaches the curve $r(s)$ as
$\varepsilon \to 0$.

Finally, we consider the function $h(s)$ defined in the Introduction, so that
each region $\Omega_\varepsilon^\alpha$  is properly deformed, and the
result is
$$\Lambda_\varepsilon^\alpha := \left\{ x \in \mathbb R^3: x = r(s) +
\varepsilon h(s) y_1 N_\alpha(s) + \varepsilon y_2 h(s) B_\alpha(s),\,
s \in  I, (y_1, y_2) \in S \right\}.$$
From now on we will omit the symbol $\alpha$ in most notations and write $dx=dsdy_1dy_2$ and~$dy=dy_1dy_2$.

In this work we study the behavior of a free quantum particle that moves in
$\Lambda_\varepsilon$,  and initially with Dirichlet boundary condition at the boundary
$\partial\Lambda_\varepsilon$.
Thus, we initially consider the family of quadratic forms
\begin{equation}\label{ddd6}
b_\varepsilon (\psi) := \int_{\Lambda_\varepsilon} |\nabla \psi|^2 dx,
\qquad
\dom b_\varepsilon = {\mathcal H}_0^1(\Lambda_\varepsilon),
\end{equation}
which is associated with the Dirichlet Laplacian operator  $-\Delta_\varepsilon$ in
$\Lambda_\varepsilon$.
The symbol $\nabla = (\partial_s, \nabla_y)$, $\nabla_y=(\partial_{y_1},\partial_{y_2})$, denotes the gradient in the  coordinates $(s,y_1,y_2)$ in~$\mathbb R^3$.

\section{Quadratic forms}\label{ffqq}

As usual in this kind of problems, in this section we perform a change of
variables so that the integration region
in \eqref{ddd6}, and consequently the domains, become independent
of~$\varepsilon > 0$. Then, for the singular limit $\varepsilon\to0$, customary ``regularizations'' will be employed.

Consider the mapping
$$\begin{array}{cccl}
f_\varepsilon: & I \times  S & \to & \Lambda_\varepsilon \\
           & (s, y_1, y_2)    & \mapsto & r(s) + \varepsilon\, h(s) \left(  y_1
N_\alpha(s) + y_2 B_\alpha(s)\right),
\end{array}$$
and suppose the boundedness $\| k\|_\infty, \|\tau\|_\infty,
\|\alpha'\|_\infty < \infty$.
These conditions are to guarantee that
$f_\varepsilon$ will be a diffeomorphism.
With this change of variables we work with a fixed region for all
$\varepsilon > 0$; more precisely, the domain of the quadratic form~\eqref{ddd6}  turns out to be ${\mathcal H}_0^1(I \times S)$.
On the other hand, the price to be paid is a nontrivial Riemannian metric $G=
G_\varepsilon^\alpha$
which is induced by $f_\varepsilon$, i.e.,
$$G=(G_{ij}),\qquad G_{ij} = \langle e_i, e_j \rangle = G_{ji},
\qquad
1 \leq i, j \leq 3,$$
where
$$e_1 = \frac{\partial f_\varepsilon}{\partial s}, \qquad
e_2 = \frac{\partial f_\varepsilon}{\partial y_1}, \qquad
e_3 = \frac{\partial f_\varepsilon}{\partial y_2}.$$

Some calculations show that in the Frenet frame
\begin{eqnarray*}
J & = &
\left(\begin{array}{c}
e_1 \\
e_2 \\
e_3
\end{array}\right)
\\
& = &
\left(\begin{array}{ccc}
\beta_\varepsilon &
-\varepsilon h (\tau+\alpha')  \langle z_\alpha^\perp, y \rangle +
\varepsilon h' \langle z_\alpha, y \rangle &
\varepsilon h (\tau+\alpha')  \langle z_\alpha, y \rangle + \varepsilon h'
\langle z_\alpha^\perp, y \rangle \\
0 & \varepsilon h \cos \alpha & \varepsilon h \sin \alpha \\
0 & -\varepsilon h \sin \alpha & \varepsilon h \cos \alpha
\end{array}\right),
\end{eqnarray*}
where
$$\beta_\varepsilon(s, y) = 1 - \varepsilon h(s) k(s) \langle z_\alpha, y
\rangle, \quad
z_\alpha : = (\cos \alpha, - \sin \alpha), \quad
z_\alpha^\perp : = (\sin \alpha, \cos \alpha).$$

The inverse matrix of~$J$ is given by
$$J^{-1} =
\left(\begin{array}{ccc}
\frac{1}{\beta_\varepsilon} &
\frac{1}{\beta_\varepsilon} \left[(\tau + \alpha') y_2 - \frac{h'}{h} y_1
\right] &
\frac{1}{\beta_\varepsilon} \left[  - (\tau + \alpha') y_1 - \frac{h'}{h}
y_2 \right] \\
0 &  \frac{\cos \alpha}{\varepsilon h} &   \frac{-\sin \alpha}{\varepsilon
h} \\
0 &  \frac{\sin \alpha}{\varepsilon h} &  \frac{\cos \alpha}{\varepsilon h}
\end{array}\right).$$

Note that  $J J^t = G$ and $\det J = | \det G|^{1/2} = \varepsilon^2
h^2(s) \beta_\varepsilon(s, y)$.
Since $k$ and  $h$ are bounded functions, for $\varepsilon$ small enough
$\beta_\varepsilon$ does not vanish in $I \times S$.  Thus,
$\beta_\varepsilon > 0$ and $f_\varepsilon$ is
a local diffeomorphism. By requiring that $f_\varepsilon$ is injective
(that is, the tube is not self-intersecting),  a global
diffeomorphism is obtained.

Introducing the notation
$$\left| \left| \psi \right| \right|_G^2
:=\int_{I \times S} | \psi(s,y) |^2 \varepsilon^2 h^2(s)
\beta_\varepsilon(s, y) \,ds dy,$$
we obtain a sequence of quadratic forms
$$\tilde{b}_\varepsilon
(\psi) : = \left| \left| J^{-1}\nabla \psi \right| \right|_G,
\qquad
\dom \tilde{b}_\varepsilon = {\mathcal H}_0^1(I \times S, G).$$
More precisely, the above change of coordinates was obtained by a
unitary transformation
$$\begin{array}{cccl}
U_\varepsilon: & \LL^2(\Lambda_\varepsilon) & \to & \LL^2(I \times S, G) \\
           & \phi   & \mapsto & \phi \circ f_\varepsilon
           \end{array}.$$
However, we still  denote $U_\varepsilon \psi$ by $\psi$.

Recall that $\lambda_0$ is the lowest eigenvalue of the negative Laplacian with Dirichlet boundary conditions in the cross section region~$S$, and $u_0\ge0$ (see Equation~\eqref{eqU0lambda0}) the corresponding eigenfunction of this restricted problem. This eigenfunction~$u_0$ is directly related to transverse oscillations in~$\Lambda_\varepsilon$.
Due to this fact, in  \cite{Bou,Cr1}  the authors have remove the diverging energy $\lambda_0 / \varepsilon^2$ from their quadratic forms.
In our case, as the boundary of the tubes were multiplied by~$h(s)$, we
subtract the terms of the
form $\lambda_0 /( \varepsilon M)^2$, i.e., since
$0 < h(s) \le M$, for all $s \in I$,
we eliminate the possible ``least transverse energy.''

Therefore, we turn to the study of the sequence of quadratic forms
$$\tilde{g}_\varepsilon(\psi)   : =  \left(
\left| \left| J^{-1} \nabla \psi \right| \right|_G^2 -
\frac{\lambda_0}{\varepsilon^2 M^2} \| \psi\|_G^2
+ c \| \psi\|_G^2 \right),$$
where $c$ is a positive constant to be chosen later on. After the norms are written out, we obtain
\begin{eqnarray*}
\tilde{g}_\varepsilon (\psi) &   =  &
\varepsilon^2 \int_{I \times S}
\Big( \frac{1}{\beta_\varepsilon^2(s,y)}
\left|\psi' +  \nabla_y \psi \cdot
R y (\tau+\alpha')(s) - \nabla_y \psi \cdot  y \,\frac{h'(s)}{h(s)}
\right|^2  \\ &+&   \frac{ |\nabla_y  \psi|^2 }{\varepsilon^2 h(s)^2}
- \frac{\lambda_0}{\varepsilon^2 M^2} |\psi|^2 + c |\psi|^2 \Big) 
h(s)^2 \beta_\varepsilon(s, y)\,ds dy\; .
\end{eqnarray*}
Note that $\dom \tilde{g}_\varepsilon = {\mathcal H}_0^1(I \times S)$ is a
subspace of
$\LL^2(I \times S,  h(s)^2 \beta_\varepsilon(s, y))$.
We observe that the factor
$|\nabla_y  \psi|^2  / (\varepsilon h(s))^2$ is directly related to 
transverse oscillations
of the particle. This term diverges as~$\varepsilon \to 0$, but we control
this fact by subtracting  $\lambda_0 /( \varepsilon M)^2
|\psi|^2$ from the quadratic form (a renormalization).

It will be convenient to work in the space~$\LL^2(I \times S,  \beta_\varepsilon(s,
y))$; so we consider the isometry
$$\begin{array}{cccl}
& \LL^2(I \times S, \beta_\varepsilon) & \to & \LL^2(I \times S, h(s)^2
\beta_\varepsilon) \\
           & v   & \mapsto & v h^{-1} 
\end{array}.$$
This change of variables and the division by the global factor~$\varepsilon^2$ (a common singular factor due to the ``change of dimension'' as $\varepsilon\to0$) leads to
\begin{eqnarray*}\hat{g}_\varepsilon (v) &: =&
\int_{I \times S} \Big( \frac{1}{\beta_\varepsilon(s,y)}
\left|v' - v \,\frac{h'(s)}{h(s)} + \nabla_y v \cdot
R y (\tau+\alpha')(s) - \nabla_y v \cdot  y \,\frac{h'(s)}{h(s)} \right|^2
 \\
 &+& \frac{\beta_\varepsilon(s,y)}{\varepsilon^2 h(s)^2} |\nabla_y 
v|^2
- \frac{\beta_\varepsilon(s, y)}{\varepsilon^2 M^2} |v|^2 + c
\beta_\varepsilon(s, y)  |v|^2 \Big)\,ds dy,
\end{eqnarray*}
with $\dom \hat{g}_\varepsilon = {\mathcal H}_0^1(I \times S)$, again as a
subspace of
$\LL^2(I \times S, \beta_\varepsilon(s, y))$.
However, this latter space can be identified with
$\LL^2(I \times S)$, for all~$\varepsilon>0$, since $\beta_\varepsilon(s,
y)$ converges uniformly to  $1$
as $\varepsilon \to 0$. Hence we introduce the form
\begin{eqnarray*}g_\varepsilon (v) &: =&
\int_{I \times S} \Big( \left|v' - v \,\frac{h'(s)}{h(s)} + \nabla_y v \cdot
R y (\tau+\alpha')(s) - \nabla_y v \cdot  y \,\frac{h'(s)}{h(s)} \right|^2
\\ &+& \frac{\beta_\varepsilon(s,y)}{\varepsilon^2 h(s)^2} |\nabla_y 
v|^2
- \frac{\beta_\varepsilon(s, y)}{\varepsilon^2 M^2} |v|^2 + c |v|^2
\Big)\,ds dy,
\end{eqnarray*}
with $\dom g_\varepsilon = {\mathcal H}_0^1(I \times S)$.

Let $\hat{G}_\varepsilon$ and  $G_\varepsilon$ be the self-adjoint
operators associated with the quadratic forms~$\hat{g}_\varepsilon$ and~$g_\varepsilon$, respectively.

\

\begin{teo}\label{teoap}
 For $\varepsilon$ small enough, there exists~$C > 0$ so that
$$\left\| \hat{G}_\varepsilon^{-1} - G_\varepsilon^{-1} \right\| \leq C \varepsilon.$$
\end{teo}

\

This theorem  follows basically from the fact that $\beta_\varepsilon (s,
y) \to 1$
uniformly as $\varepsilon \to 0$. Its proof is presented in the Appendix.

Due to the above changes of variables and Theorem~\ref{teoap},  we may consider the
sequence of quadratic forms~$g_\varepsilon$ in what follows.

\section{Reduction of dimension}\label{redim}

Recall that
$u_0(y)$ is the positive and normalized eigenfunction corresponding to the
first eigenvalue $\lambda_0$ of the
Laplacian in ${\mathcal H}_0^1(S)$.
After the orthogonal decomposition  $\LL^2(\mathbb R \times S) = {\mathcal L}
\oplus {\mathcal L}^\perp,$
for $\psi \in \LL^2(\mathbb R \times S)$, we can write
$$\psi(s, y) = w(s) u_0(y) + \eta(s, y),$$
with $w \in \LL^2(I)$ and $\eta \in {\mathcal L}^\perp$.
We observe that  $\eta\in {\mathcal L}^\perp$ implies
$$\int_S u_0(y) \eta(s, y) dy = 0, \qquad
\hbox{a.e.}[s].$$

Note that  $wu_0 \in {\mathcal H}_0^1(I \times S)$ if $w \in {\mathcal
H}_0^1(I)$.
For $\psi \in {\mathcal H}_0^1(\mathbb R \times S)$, write
$\psi = w u_0 + \eta$ with $w \in {\mathcal H}_0^1(I)$ and $\eta \in {\mathcal
H}_0^1(\mathbb R \times S) \cap {\mathcal L}^\perp$.

First we study the quadratic form  $g_\varepsilon$ restricted to
the subspace
\mbox{${\mathcal H}_0^1(I \times S) \cap {\mathcal L}$.}
For $w \in {\mathcal H}_0^1(I)$, some calculations show that
\begin{eqnarray*}
g_\varepsilon( w u_0) = \int_{I} \left[
|w'|^2 + \vartheta(s) |w|^2  +
\zeta_\varepsilon(s, y)
\left(\frac{\lambda_0}{\varepsilon^2h^2(s)}-\frac{\lambda_0}{\varepsilon^2M^2}\right)
|w|^2 + c |w|^2 \right] ds,
\end{eqnarray*}
where
$$\vartheta(s) = C_1(S) (\tau(s) + \alpha'(s))^2
+ (C_2(S) - 1) \left(\frac{h'(s)}{h(s)}\right)^2 - 2 C_3(S)
(\tau(s)+\alpha'(s)) \,\frac{h'(s)}{h(s)}$$
and
$$\zeta_\varepsilon(s, y) =  1 - \varepsilon\, k(s) h(s) \left\langle
z_\alpha(s), F(S) \right\rangle .$$
The constants  $C_1(S)$, $C_2(S)$ and $C_3(S)$ that appear in the
definition of~$\vartheta$ depend only on the region $S$
and are explicitly given by
$$C_1(S) = \int_S |\langle \nabla_y u_0, R y \rangle|^2 dy, \qquad
C_2(S) = \int_S |\langle \nabla_y u_0, y \rangle|^2 dy,$$ and
$$C_3(S) = \int_S \langle \nabla_y u_0, R y \rangle \langle \nabla_y u_0, 
y \rangle dy.$$
The vector $F(S) = (F_1(S), F_2(S))$ in the definition of
$\zeta_\varepsilon$ also depends only on the
region~$S$, and its components are given by
$$F_1(S) = \int_S y_1 |u_0|^2 dy\qquad \hbox{and}\qquad
F_2(S) =  \int_S y_2 |u_0|^2 dy.$$
Under such restrictions,  the quadratic form $b_\varepsilon$ in ${\mathcal H}_0^1(I)$ can be written in terms of the form $t_\varepsilon=t_{\varepsilon,c}$ given by
\begin{equation}\label{eqdefteps}
t_\varepsilon(w) : = g_\varepsilon(wu_0)=\int_I \left( |w'|^2 + W_\varepsilon(s) |w|^2
\right) ds,
\end{equation}
with
\begin{equation}\label{aaa1}
W_\varepsilon(s)  : =  \vartheta(s)+ c + \zeta_\varepsilon(s, y)
\left(\frac{\lambda_0}{\varepsilon^2 h^2(s)}-\frac{\lambda_0}{
\varepsilon^2 M^2}\right).
\end{equation}
We choose the constant~$c$ so that $c > \|v\|_\infty + (1 / M^2)\|
k(s)^2 / 4\|_\infty$.

Since $k(s)$ and $h(s)$ are bounded functions, there exist
$\varepsilon_1 > 0$  and $\delta > 0$ so that, for all $s\in I$,  
$$ 
1 - \varepsilon\, k(s) h(s) \langle z_\alpha(s), F(S) \rangle > \delta \qquad \hbox{and} \qquad
1 - \varepsilon\, k(s) h(s) \langle z_\alpha(s), y \rangle > \delta,
$$
for all $\varepsilon < \varepsilon_1$.
In what follows, we tacitly assume that  $\varepsilon < \varepsilon_1$.

The self-adjoint operator associated with $t_\varepsilon$  in $\LL^2(I)$ is
$$(T_{\varepsilon,c} w)(s) := - w''(s) + W_\varepsilon(s) w(s), \qquad
\dom{T_{\varepsilon,c}} = {\mathcal H}^2(I)\cap {\mathcal H}_0^1(I).$$

From now on we denote by $-\Delta_{\varepsilon,c}$ the operator
$-\Delta_\varepsilon + c\Id$ and write $T_\varepsilon=T_{\varepsilon,c}-c\Id$.
Next we discuss how the resolvent operator
$\left(- \Delta_{\varepsilon,c} - \lambda_0 / \varepsilon^2 M^2 \Id \right)^{-1}$
can be approximated by $T_{\varepsilon,c}^{-1}\oplus 0$, where $0$  is the
null operator on the
subspace~${\mathcal L}^\perp$. Such result gives a quantitative indication of how $-\Delta_\varepsilon$ is approximated by~$T_\varepsilon$.

\

\begin{lem}\label{ddd10}
{\rm Suppose that  $I$ is  a bounded interval. Then, there exists
$C_6 > 0$ so that
$$t_\varepsilon(w) \geq C_6^{-1}\, \varepsilon^{-1} \int_I |w|^2 ds,
\qquad \forall w\in {\mathcal H}_0^1(I),\;0 < \varepsilon < \varepsilon_1.$$
}
\end{lem}

\

By noting  that
$$ \frac{\varepsilon^2 W_\varepsilon (s)}{s^2} \geq \frac{\lambda_0}{s^2}
\delta \left(
\frac{1}{h(s)^2} - \frac{1}{M^2} \right),$$
the proof of Lemma~\ref{ddd10} is similar to the proof of Lemma $2.1$
in~\cite{Frie}, and it will not be reproduced here.

By following \cite{Bou}, for each  $\xi \in \mathbb R^2$, we consider the
following perturbed problem
$$-{\rm div}[(1 - (\xi \cdot y )) \nabla_y u] =
\lambda (1 - (\xi \cdot y)) u, \qquad
u \in {\mathcal H}_0^1(S).$$

By taking $\xi= \varepsilon h(s) k(s) z_\alpha$, for  $\varepsilon$
small enough, the perturbed  operator
is positive and with compact resolvent. Denote by $\lambda(\xi) > 0$
its first eigenvalue, i.e.,
$$\lambda(\xi) =
\inf_{\{u \in {\mathcal H}_0^1(S):\: u \neq 0\}}
\frac{\int_S (1 - (\xi \cdot  y )) |\nabla_y u|^2 dy}{\int_S (1 - (\xi
\cdot  y )) |u|^2 dy}.$$
Thus, for $v \in {\mathcal H}_0^1(\mathbb R \times S)$,
\begin{equation}\label{ddd8}
\frac{1}{\varepsilon^2} \int_S \beta_\varepsilon(s,y)
( | \nabla_y v|^2 - \lambda_0 |v|^2) dy \geq \gamma_\varepsilon(s) \int_S
\beta_\varepsilon(s,y) |v|^2 dy
\qquad
{\rm a.e.}[s],
\end{equation}
where
$$\gamma_\varepsilon(s) := \frac{\lambda(\varepsilon h(s) k(s)
z_{\alpha}(s)) - \lambda_0}{\varepsilon^2}.$$

Using the fact that $h(s)$ and $k(s)$ are bounded functions, it is possible to prove
that
$\gamma_\varepsilon(s)$ converges uniformly as $\varepsilon \to 0$ to a bounded function (see
Proposition $4.1$ in \cite{Bou}). This will be used in the proof of Lemma~\ref{eee4}.

\begin{lem}\label{eee4}
{\rm
Let $I$ denote either $\mathbb R$ or a bounded interval. Then, for $\eta \in 
{\mathcal H}_0^1(I \times S) \cap {\mathcal L}^\perp$, there exists  $C_7\in\R$ so
that,
for $\varepsilon$ small enough,
$$g_\varepsilon(\eta) \geq  \frac{C_7}{\varepsilon^2 M^2} \| \eta \|^2.$$}
\end{lem}
\begin{dm}
Let  $\lambda_1$ be the second eigenvalue of the Laplacian in ${\mathcal
H}_0^1(S)$, and pick
$\eta \in {\mathcal H}_0^1(\mathbb R \times S) \cap {\mathcal L}^\perp$.

Since
$h(s) \leq M$, for all $s \in I$, we have
$$\int_S
\beta_\varepsilon(s, y)   \left(
\frac{|\nabla_y \eta|^2}{\varepsilon^2 h(s)^2}
- \lambda_1 \frac{|\eta|^2}{\varepsilon^2 M^2}  \right)dy
\geq
\int_S \beta_\varepsilon(s, y)  \left(
\frac{|\nabla_y \eta|^2}{\varepsilon^2 M^2}
- \lambda_1 \frac{|\eta|^2}{\varepsilon^2 M^2}  \right)dy.$$
By \eqref{ddd8}, it follows that
$$\int_S \beta_\varepsilon(s, y)  \left(
\frac{|\nabla_y \eta|^2}{\varepsilon^2 M^2}
- \lambda_1 \frac{|\eta|^2}{\varepsilon^2 M^2}  \right)dy
\geq
\frac{\gamma_\varepsilon(s)}{M^2} \int_S \beta_\varepsilon(s, y) |\eta|^2 dy.$$

Since $\gamma_\varepsilon(s)$ converges  uniformly as  $\varepsilon \to 0$,
there exists $C_8 \in \mathbb R$ so that, for $\varepsilon$ small enough,
$$\frac{\gamma_\varepsilon(s)}{M^2} \geq C_8, \qquad \forall s \in I.$$
Thus,
$$\frac{\gamma_\varepsilon(s)}{M^2} \int_S \beta_\varepsilon(s, y) |\eta|^2 dy
\geq
C_8 \int_S  \beta_\varepsilon(s, y) |\eta|^2 dy,$$
and so
$$\int_{I \times S}
\beta_\varepsilon(s, y)   \left(
\frac{|\nabla_y \eta|^2}{\varepsilon^2 h(s)^2}
- \lambda_1 \frac{|\eta|^2}{\varepsilon^2 M^2}  \right)dy ds
\geq
C_8 \int_{I \times S}  \beta_\varepsilon(s, y) |\eta|^2 dy ds.$$

Adding and subtracting the term  $  \frac{\lambda_0}{\varepsilon^2 M^2}
\int_{I \times S} \beta_\varepsilon (s, y) |\eta|^2 dyds$
on the left hand side of the above inequality, we obtain
\begin{eqnarray*}
\int_{I \times S}
&\beta_\varepsilon(s, y)& \left(
\frac{|\nabla_y \eta|^2}{\varepsilon^2 h(s)^2}
- \lambda_0  \frac{|\eta|^2}{\varepsilon^2 M^2}  \right) dy ds\\
& \geq &
C_8 \int_{I \times S} \beta_\varepsilon(s, y)  |\eta|^2 dy +
\frac{(\lambda_1 - \lambda_0)}{\varepsilon^2 M^2} \int_{I\times S}
\beta_\varepsilon(s, y) |\eta|^2 dy ds.
\end{eqnarray*}

Now, for  $\varepsilon$ small enough, there exists~$C_9$ so that
\begin{eqnarray*}
g_\varepsilon(\eta) &\geq &
\int_{I \times S}
\beta_\varepsilon(s, y) \left(
\frac{|\nabla_y \eta|^2}{\varepsilon^2 h(s)^2}
- \lambda_0 \frac{|\eta|^2 }{\varepsilon^2 M^2} \right) dy ds + c \int_{I
\times S} |\eta|^2  dy ds\\
& \geq &
C_8 \delta \int_{I \times S}  |\eta|^2 dy ds +
\frac{(\lambda_1 - \lambda_0)}{\varepsilon^2 M^2} \int_{I \times S}
\beta_\varepsilon(s, y) |\eta|^2 dy ds
+ c \int_{I \times S}  |\eta|^2 dy ds\\
& \geq &
\frac{C_9}{\varepsilon^2 M^2} \int_{I \times S} \beta_\varepsilon(s, y)
|\eta|^2 dy ds\\
& \geq &
\frac{C_9}{\varepsilon^2 M^2} \delta \int_{I \times S} |\eta|^2 dy ds.
\end{eqnarray*}
Finally, it is enough to take $C_7= C_9\, \delta$ to complete the proof of the
lemma.
$\cq$
\end{dm}

Now we are ready to state and prove the main result of this section; it will rest on results presented in Section~$3$ of~\cite{Sol}, combined with the previous lemmas.

\begin{teo}\label{aaa4}
{\rm Let $I$ denote either~$\R$ or a bounded interval. Then there exists~$C_{10} > 0$ so that, for $\varepsilon$ small enough,
$$\left|\left| \left(- \Delta_{\varepsilon,c} -
\frac{\lambda_0}{\varepsilon^2 M^2} \Id \right)^{-1}
- \left(T_{\varepsilon,c}^{-1} \oplus 0\right) \right| \right| \leq C_{10}\,
\varepsilon^{3/2},$$
where $0$ denotes the null operator on the subspace ${\mathcal L}^\perp$.
}
\end{teo}
\begin{dm}{\rm
For $\psi \in {\mathcal H}_0^1(\mathbb R \times S)$ write
$$\psi(s,y) = w(s) u_0(y) + \eta(s, y),$$
with $w \in {\mathcal H}_0^1(I)$ and  $\eta \in {\mathcal H}_0^1(\mathbb R
\times S) \cap {\mathcal L}^\perp$. Thus,
the quadratic form $g_\varepsilon(\psi)$ can be rewritten as
$$g_\varepsilon(\psi) = t_\varepsilon(w) + g_\varepsilon(\eta) + 2 
m_\varepsilon(wu_0, \eta),$$
where  $t_\varepsilon(w) = g_\varepsilon(wu_0)$ (see \eqref{eqdefteps}) and
\begin{eqnarray*}
m_\varepsilon (wu_0, \eta) & = &
\int_{I \times S} dy ds \Big[
\left( w' u_0 - w u_0 \,\frac{h'}{h} + w \nabla_y u_0 \cdot Ry
(\tau+\alpha') - w \nabla_y u_0 \cdot y \,\frac{h'}{h} \right) \\
& \times &
\left(\eta' - \eta \,\frac{h'}{h}  +  \nabla_y \eta \cdot Ry (\tau+\alpha') -
\nabla_y \eta \cdot y \,\frac{h'}{h} \right) \Big]\\
& - &
\int_{I \times S} dy ds\;
k(s) h(s) \langle z_\alpha, y \rangle w \left(\frac{\nabla_y u_0 \nabla_y
\eta}{\varepsilon h^2} -
\lambda_0 \frac{u_0 \eta}{\varepsilon M^2}  \right).
\end{eqnarray*}

We are going to show that  $t_\varepsilon(w)$, $g_\varepsilon(\eta)$ and
$m_\varepsilon(wu_0, \eta)$ satisfy the conditions
(3.2), (3.3), (3.4) and (3.5) in Section~$3$ of \cite{Sol},
and so the theorem will follow. Conditions (3.2), (3.3) and (3.4) are obtained by applying Lemmas~\ref{ddd10} and~\ref{eee4} above. We need only to verify  condition~(3.5), i.e., that there exists a
function $q(\varepsilon)$ so that for each
$\psi \in {\mathcal H}_0^1(I \times S)$
\begin{equation}\label{eee2}
|m_\varepsilon(wu_0, \eta)|^2 \,\leq\, q(\varepsilon)^2\, t_\varepsilon(w)
\,g_\varepsilon(\eta), \qquad
q(\varepsilon) \to 0 \qquad (\varepsilon \to 0).
\end{equation}

We write
$$m_\varepsilon(wu_0, \eta) = m_\varepsilon^1(wu_0, \eta) -
m_\varepsilon^2(wu_0, \eta) + m_\varepsilon^3(wu_0, \eta) -
m_\varepsilon^4(wu_0, \eta)
- m_\varepsilon^5(wu_0, \eta),$$
where
\begin{eqnarray*}
m_\varepsilon^1(w u_0, \eta)  & : = &
\int_{\mathbb R \times S}
w' u_0 \left( \eta' - \eta \,\frac{h'}{h} + \nabla_y \eta \cdot R y (\tau + \alpha')
- \nabla_y \eta \cdot y \,\frac{h'}{h} \right) \,ds dy\, , \\
m_\varepsilon^2(wu_0, \eta) & : = &
\int_{\mathbb R \times S}
w u_0 \,\frac{h'}{h} \left(  \eta' - \eta \,\frac{h'}{h} + \nabla_y \eta \cdot R y
(\tau + \alpha')
- \nabla_y \eta \cdot y \,\frac{h'}{h} \right) \,ds dy\, , \\
m_\varepsilon^3(wu_0, \eta) & : = & \int_{\mathbb R \times S}
w \nabla_y u_0 \cdot Ry (\tau+\alpha')
\left(  \eta' - \eta \,\frac{h'}{h} + \nabla_y \eta \cdot R y (\tau + \alpha')
- \nabla_y \eta \cdot y \,\frac{h'}{h} \right) dy ds\, , \\
m_\varepsilon^4(wu_0, \eta) & := &
\int_{\mathbb R \times S}
w \nabla_y u_0 \cdot y \,\frac{h'}{h}
\left(  \eta' - \eta \,\frac{h'}{h} + \nabla_y \eta \cdot R y (\tau + \alpha')
- \nabla_y \eta \cdot y \,\frac{h'}{h} \right) \,ds dy\, , \\
m_\varepsilon^5(wu_0, \eta) & : = &
\int_{I \times S}
\frac{k(s) h(s) \langle z_{\alpha} , y \rangle}{\varepsilon}
w \left(\frac{\nabla_y u_0 \nabla_y \eta} {h^2} - \lambda_0
\frac{u_0 \eta}{M^2} \right) dy ds.
\end{eqnarray*}

Now we are going to estimate each of the above terms. Let
$$  H_1: = \left| \left| \frac{h'}{h} \right| \right|_\infty,
\qquad
H_2 : = \left| \left| \tau + \alpha ' \right| \right|_\infty,$$
and recall that
$$C_1(S) = \int_S |\langle \nabla_y u_0, R y \rangle|^2 dy \qquad
\hbox{and} \qquad
C_2(S) = \int_S |\langle \nabla_y u_0, y \rangle|^2 dy.$$

By Green identities and some calculations, we get
$$\int_S u_0 \langle \nabla_y \eta, Ry \rangle dy = - \int_S \langle \nabla_y
u_0, Ry \rangle \eta dy,$$
$$\int_S u_0 \langle \nabla_y \eta, y \rangle dy = - \int_S  \langle \nabla_y
u_0, y \rangle  \eta dy.$$
Hence:
\begin{eqnarray*}
\bullet \quad
|m_\varepsilon^1(wu_0, \eta)| & \leq &
H_2^{1/2} C_1(S) \left( \int_I |w'|^2 ds \right)^{1/2} \left( \int_{I
\times S}  |\eta|^2 dy ds\right)^{1/2} \\
& + &
H_1^{1/2} C_2(S) \left( \int_I |w'|^2 ds \right)^{1/2} \left( \int_{I
\times S}  |\eta|^2 dy ds \right)^{1/2} \\
& \leq &
C_{11}\, \varepsilon\,  t_\varepsilon(w)^{1/2}  g_\varepsilon(\eta)^{1/2}; \\
\end{eqnarray*}

\begin{eqnarray*}
\bullet \quad  | m_\varepsilon^2(wu_0, \eta)| & \leq &
\left(\int_{I \times S} |w|^2 |u_0|^2 \left(\frac{h'}{h}\right)^2 dy
ds\right)^{1/2} g_\varepsilon(\eta)^{1/2}\\
& \leq &
H_1 \left(\int_I |w|^2 ds\right)^{1/2} g_\varepsilon(\eta)^{1/2}\\
& \leq &
\frac{H_1}{C_6^{1/2}}\, \varepsilon^{1/2}\, t_\varepsilon(w)^{1/2}
g_\varepsilon(\eta)^{1/2};\\
\end{eqnarray*}

\begin{eqnarray*}
\bullet \quad |m_\varepsilon^3(wu_0, \eta)| & \leq &
\left(\int_{I \times S} |w|^2 |\langle \nabla_y u_0,  Ry \rangle|^2
(\tau+\alpha')^2 dy ds \right)^{1/2}
g_\varepsilon(\eta)^{1/2} \\
& = &
\left(\int_I |w|^2 C_1(S) (\tau+\alpha')^2 ds \right)^{1/2}
g_\varepsilon(\eta)^{1/2}\\
& \leq &
C_1(S)^{1/2} H_2 \left(\int_I |w|^2 ds\right)^{1/2} g_\varepsilon(\eta)^{1/2} \\
& \leq &
C_1(S)^{1/2}  \frac{H_2}{C_6^{1/2}}\, \varepsilon^{1/2}\,
t_\varepsilon(w)^{1/2} g_\varepsilon(\eta)^{1/2};\\
\end{eqnarray*}

\begin{eqnarray*}
\bullet \quad
|m_\varepsilon^4(wu_0, \eta)| & \leq &
\left(\int_{I \times S} |w|^2 |\langle \nabla_y u_0,  y \rangle |^2
\left(\frac{h'}{h}\right)^2 dy ds \right)^{1/2} g_\varepsilon(\eta)^{1/2} \\
& = &
\left(\int_I |w|^2 C_2(S) \left(\frac{h'}{h}\right)^2 ds \right)^{1/2}
g_\varepsilon(\eta)^{1/2}\\
& \leq &
C_2(S)^{1/2} H_1 \left(\int_I |w|^2 ds\right)^{1/2} g_\varepsilon(\eta)^{1/2} \\
& \leq &
C_2(S)^{1/2}  \frac{H_1}{C_6^{1/2}}\, \varepsilon^{1/2}\,
t_\varepsilon(w)^{1/2} g_\varepsilon(\eta)^{1/2}.
\end{eqnarray*}

Additional calculations show that
$$\int_{S} y_1 \langle \nabla_y u_0, \nabla_y \eta \rangle  =
- \int_{S} f_1(y) \eta  dy ds,$$
$$\int_{S} y_2 \langle \nabla_y u_0 \nabla_y \eta \rangle =
- \int_{S} f_2(y) \eta dy ds,$$
where
$$f_1(y) : = \left(\frac{\partial u_0}{\partial y_1} + y_1 \frac{\partial^2
u_0}{\partial y_1^2}
+ y_1 \frac{\partial^2 u_0}{\partial y_2^2} \right),$$
$$f_2(y) : = \left(\frac{\partial u_0}{\partial y_2} + y_2 \frac{\partial^2
u_0}{\partial y_2^2}
+ y_2 \frac{\partial^2 u_0}{\partial y_1^2} \right).$$
Thus, there exist $C_{12}$ and $C_{13}$ so that
\begin{eqnarray*}
|m_\varepsilon^5(wu_0, \eta)| & \leq  &
\left|
\int_{I \times S} k(s) \cos \alpha(s) y_1  w  \frac{\nabla_y u_0 \nabla_y
\eta}{\varepsilon h} dy ds \right| \\
& + & \left| \int_{I \times S} k(s) h(s) \cos \alpha(s) y_1 \lambda_0 w
\frac{u_0 \eta}{\varepsilon M^2}   dy ds \right| \\
& + &
\left|
\int_{I \times S} k(s) \sin \alpha(s) y_2  w  \frac{\nabla_y u_0 \nabla_y
\eta}{\varepsilon h} dy ds \right| \\
& + & \left| \int_{I \times S} k(s) h(s) \sin \alpha(s) y_2 \lambda_0 w
\frac{u_0 \eta}{\varepsilon M^2}   dy ds \right| \\
& \leq &
\frac{C_{12}}{\varepsilon} \left(\int_\mathbb R |w|^2 ds\right)^{1/2}
\left(\int_{\mathbb R \times S} |\eta|^2 dy ds\right)^{1/2}\\
& \leq &
C_{13}\, \varepsilon^{1/2}\, t_\varepsilon(w)^{1/2} g_\varepsilon(\eta)^{1/2}.
\end{eqnarray*}

By the above estimates it follows that there exists
$C_{14} > 0$ so that
$$|m_\varepsilon(wu_0, \eta)|^2 \leq C_{14}\, \varepsilon\, t_\varepsilon(w)\,
g_\varepsilon(\eta),$$
and so \eqref{eee2} is proven.
By applying Proposition~$3.1$ of \cite{Sol}, it is found that there exists
$C_{10}$ so that, for $\varepsilon$ small enough,
$$\left| \left| \left(- \Delta_{\varepsilon,c} - \frac{\lambda_0}{\varepsilon^2
M^2} \Id \right)^{-1}
- \left(T_{\varepsilon,c}^{-1} \oplus 0\right) \right| \right| \leq C_{10}\,
\varepsilon^{3/2}.$$
The proof of the theorem is complete. $\cq$
}
\end{dm}

\section{Bounded interval and  Dirichlet condition}\label{qqq3}

In this section we  suppose  that $I=[-a,b]$ is a bounded interval and  the condition at
the boundary $\partial \Lambda_\varepsilon$ is
Dirichlet.
Since $I$ is bounded, the spectrum of
$-\Delta_{\varepsilon,c}$ in $\Lambda_\varepsilon$ is purely discrete and we
denote its eigenvalues by
$l_j^c(\varepsilon)$.

The  main result in this section, that is,
Theorem~\ref{aaa3}, is a version of Theorem~\ref{mainTeor} in this context.

\begin{teo}\label{aaa3}
The limits
\begin{equation}\label{eee3new}
\mu_j = \lim_{\varepsilon \to 0}\, \varepsilon\left( l_j^c(\varepsilon) -
\frac{\lambda_0}{\varepsilon^2 M^2} \right)
\end{equation}
exist, where $\mu_j$ are the eigenvalues of a self-adjoint operator~$T$ (see Equation~\eqref{eqweo}) acting in
$\LL^2(\mathbb R)$.
\end{teo}

To prove this theorem we need  some previous results; we will
follow~\cite{Frie}. Introduce the family of segments
$$I_\varepsilon = (-a \varepsilon^{-1/2}, b \varepsilon^{-1/2}),
\qquad\, \varepsilon > 0,$$ and  the family of
unitary operators $J_\varepsilon: \LL^2(I) \to \LL^2 (I_\varepsilon)$
generated by the dilation
$s \mapsto s \varepsilon^{1/2}$, that is,
\[
(J_\varepsilon\psi)(s) = \varepsilon^{1/4}\, \psi(\varepsilon^{1/2}s),
\]and identify
$\LL^2(I_\varepsilon)$ with the subspace
$$\left\{u \in \LL^2(\mathbb R): u(s) = 0\, \hbox{ a.e. in }\, \mathbb R \backslash
I_\varepsilon\right \}.$$
Set
\begin{equation}\label{aaa7}
\hat{T}_{\varepsilon,c} : = \varepsilon J_\varepsilon T_{\varepsilon,c}
J_\varepsilon^{-1},
\end{equation}
which is a self-adjoint operator acting in $\LL^2(I_\varepsilon)$.

\

\begin{teo}\label{aaa2}
{\rm In case $I=[-a,b]$ is a bounded interval, one has
$$\left| \left| \hat{T}_{\varepsilon,c}^{-1} \oplus 0 - T^{-1} \right| \right|
\to 0, \qquad \hbox{as}
\qquad
\varepsilon \to 0,$$
where $0$ is the null operator on the subspace  $\LL^2(\mathbb R
\backslash I_\varepsilon)$.}
\end{teo}

\

We have $\varepsilon J_\varepsilon W_\varepsilon(s)  J_\varepsilon^{-1} = \varepsilon W_\varepsilon(\varepsilon^{1/2} s)$, and a direct calculation shows that
$$\varepsilon J_\varepsilon W_\varepsilon(s)  J_\varepsilon^{-1} = \zeta_\varepsilon(\varepsilon^{1/2} s, y) \lambda_0 \left[ 
 M^{-3} s^2
+  \rho (\varepsilon^{1/2} s) s^3
\varepsilon^{1/2} \right] + \varepsilon\, \vartheta(\varepsilon^{1/2} s) +
\varepsilon c,$$
with $\rho \in \LL^\infty(I)$.
Since
$\zeta_\varepsilon(\varepsilon^{1/2} s, y) \to 1$ uniformly as 
$\varepsilon \to 0$,
the proof of Theorem~\ref{aaa2} is similar to the proof of
Theorem~$1.3$ in \cite{Frie}, and so it will not be repeated here.

\

\noindent{\bf Proof of Theorem~\ref{aaa3}:}
Let $l_j(T_{\varepsilon,c})$, $l_j(\hat{T}_{\varepsilon,c})$ denote the
eigenvalues of
$T_{\varepsilon,c}$ and $\hat{T}_{\varepsilon,c}$ respectively. 
Let $\psi_{j, \varepsilon}^c$  denote the eigenfunction associated with eigenvalue $l_j^c(\varepsilon)$
of $-\Delta_{\varepsilon, c}$.
Thus, there exist functions $w_{\varepsilon, c} \in \LL^2(I)$ and $U \in \cal L $ so that
$\psi_{j, \varepsilon}^c = w_{j, \varepsilon}^c u_0 + U$. Since $\mathcal L$ is invariant under
$(-\Delta_{\varepsilon, c} - \lambda_0 / \varepsilon^2 M^2 \Id)$, it 
follows that $w_{j, \varepsilon}^c u_0$ is the eigenfunction associated with the
 eigenvalue $l_j(T_{\varepsilon, c})$.
Observe also that the nonzero
eigenvalues of
$T_{\varepsilon,c}^{-1} \oplus 0$ are exactly the eigenvalues of
$T_{\varepsilon,c}^{-1}$.
Hence, by Theorem~\ref{aaa4}, we have
\begin{eqnarray*}
\left| \left( l_j^c(\varepsilon) - \frac{\lambda_0}{\varepsilon^2 M^2}
\right)^{-1} - l_j^{-1}(T_{\varepsilon,c}) \right|
& \leq &
\left| \left| \left( -\Delta_{\varepsilon,c} -
\frac{\lambda_0}{\varepsilon^2 M^2}\Id  \right)^{-1} -
T_{\varepsilon,c}^{-1} \oplus 0 \right| \right| \\
& \leq &
C_{10}\, \varepsilon^{3/2}.
\end{eqnarray*}
Thus,
$$\left| \frac{1}{\varepsilon} \left( l_j^c(\varepsilon) -
\frac{\lambda_0}{\varepsilon^2 M^2} \right)^{-1}-
\frac{1}{\varepsilon l_j^{-1}(T_{\varepsilon,c})} \right| \leq C_{10}\,
\varepsilon^{1/2}.$$
Since $l_j(\hat{T}_{\varepsilon,c}) = \varepsilon l_j(T_{\varepsilon,c})$, by
Theorem~\ref{aaa2}, we find
$$\varepsilon l_j(T_{\varepsilon,c}) \to \mu_j, \quad\varepsilon \to 0,
$$
and~\eqref{eee3new} follows.
$\cq$

\section{The Neumann case}\label{nnnn}
Here we again consider  that $I=[-a,b]$ is a bounded interval, but the Dirichlet
condition at the vertical part of
the boundary  $\partial(I \times S)$, that is, $\{(-a)\times S\cup b\times S\}$,
is replaced by Neumann condition.
Our point is that the conclusions of Theorem~\ref{aaa3} also hold true in
this case.
Although in our case the curvature and torsion can be nontrivial, the
proof in this case are similar to the proof of
Theorem~\ref{aaa3} above (and taking into account~\cite{Sol}); for this reason,
details will not be presented.

\section{The case $I = \mathbb R$ and Dirichlet condition}\label{iiii}

In this section we study the case  $I = \mathbb R$. First we give sufficient conditions for a nonempty discrete spectrum of the Dirichlet Laplacian, and then discuss the WEO and eigenvalue approximations.

\subsection{The discrete spectrum}\label{udah}

Now the spectrum of the Laplacian $- \Delta_{\varepsilon,c}$ in
$\Lambda_\varepsilon$ is not
necessarily discrete, but in this section  we will see that the essential spectrum $\sigma_{\mathrm{ess}}(- \Delta_{\varepsilon,c})$  depends on the behavior of the curvature at infinity; it will then follow that if $k(s)\to 0$ as $|s|\to\infty$, then the discrete spectrum of $- \Delta_{\varepsilon,c}$ is nonempty for $\varepsilon$ small enough.

Denote $\nu(\varepsilon) := \inf \sigma_{{\rm ess}}
(-\Delta_{\varepsilon,c})$ and let
$l_j^c(\varepsilon)$ be the eigenvalues of $-\Delta_{\varepsilon,c}$ (recall
the Dirichlet boundary condition).

\begin{teo}
If $I=\mathbb R$ and the curvature satisfies
\begin{equation}\label{eee5}
\lim_{|s| \to \infty} k(s) =0,
\end{equation}
then $\nu(\varepsilon) \to \infty$ as $\varepsilon \to 0.$
\end{teo}
\begin{dm}{\rm
Let $N : =   \limsup_{|s| \to \infty} h(s) < M$ and $\hat{I} = [-a, a]$
and define
$$\Omega_{a, \varepsilon} =\left \{(s, y): s \in \hat{I} \right\} \qquad
\hbox{and} \qquad
\Omega_{a, \varepsilon}' = \left\{(s, y): s \notin \hat{I}\right \}.$$

Let $-\Delta_{a, \varepsilon, D}^c$, $-\Delta_{a, \varepsilon, D}^{'c}$ be
the Dirichlet  Laplacian in
$\Omega_{a, \varepsilon}$ and $\Omega_{a, \varepsilon}'$ respectively.
Similarly, let
$-\Delta_{a, \varepsilon, DN}^c$, $-\Delta_{a, \varepsilon, DN}^{'c}$ be
the above Laplacian operators but with Neumann condition  at the
vertical part of the boundaries of~$\Omega_{\alpha,\varepsilon}$ and~$\Omega_{\alpha,\varepsilon}^{'}$, respectively.
Note that
\begin{equation}\label{eee6}
- \Delta_{a, \varepsilon, DN}^c + \left(- \Delta_{a, \varepsilon,
DN}^{'c}\right)
< - \Delta_{\varepsilon, c} <
- \Delta_{a, \varepsilon, D}^c + \left(- \Delta_{a, \varepsilon,
D}^{'c}\right).
\end{equation}
Therefore
$\inf \sigma_{{\rm ess}} (-\Delta_{\varepsilon, c}) \geq \inf \sigma_{{\rm
ess}} (- \Delta_{a, \varepsilon, DN}^{'c})$.

Let $q_{a, \varepsilon, DN}'$ the quadratic form associated with the
operator  $-\Delta_{a, \varepsilon, DN}^{'c}$.
Write $K_\varepsilon= \sup_{(s,y) \in \mathbb R \times S}
\beta_\varepsilon(s, y)$; we have
\begin{eqnarray*}
q_{a, \varepsilon, DN}' (\psi)
& \geq &
\left(
\inf_{(s,y) \in \mathbb R \times S}{\frac{\beta_\varepsilon(s,
y)}{\varepsilon^2 h(s)}} \right) \int_{(\mathbb R \backslash \hat{I})
\times S}
|\nabla_y \psi|^2 dy ds \\
& \geq &
\lambda_0 \left(
\inf_{(s,y) \in \mathbb R \times S}{\frac{\beta_\varepsilon(s,
y)}{\varepsilon^2 h(s)}} \right) \int_{(\mathbb R \backslash
\hat{I})\times S}
|\psi|^2 dy ds \\
& \geq &
\lambda_0
\left(\inf_{(s,y) \in \mathbb R \times S}{\frac{\beta_\varepsilon(s,
y)}{\varepsilon^2 h(s)}} \right)
\frac{1}{K_\varepsilon}
\int_{(\mathbb R \backslash \hat{I})\times S}
\beta_\varepsilon(s, y) |\psi|^2 dy ds,
\end{eqnarray*}
for all $\psi \in \dom q_{a, \varepsilon, DN}'$.
Since $k$ satisfies \eqref{eee5},
it follows that the essential spectrum of $-\Delta_{a, \varepsilon, DN}^{'c}$
is estimated from below by  $\lambda_0$ times a function that converges to
$  \frac{1}{\varepsilon^2  N}$ as $a \to \infty$.
Since the essential spectrum is a closed  subset, it follows that
$\nu(\varepsilon) \geq   \frac{\lambda_0}{\varepsilon^2 N^2}$ and
consequently
$\nu(\varepsilon) \to \infty$ as $\varepsilon \to 0$.
$\cq$
}
\end{dm}

We conclude that, under condition~\eqref{eee5},  for $\varepsilon$  small enough the discrete spectrum of
$-\Delta_{\varepsilon, c}$
is nonempty. We again stress that we have got another property that does
not depend on important geometric features of the tube.
Also the WEO~$T$ (see also Subsection~\ref{subsecEo}),
which weakly describes the asymptotic behaviors of the eigenvalues of
$-\Delta_{\varepsilon, c}$ in the sense of~\eqref{eee3new},  is not influenced by such geometric features.

\

\subsection{Weakly effective operator}\label{subsecEo}

The goal of this section is to show that Theorems~\ref{aaa4}, \ref{aaa3} 
and~\ref{aaa2} have a similar counterpart in case~$I = \mathbb R$.
In \cite{Sol} these theorems are proven for two dimensional
strips, and here we argue that those proofs can be adapted to our three
dimensional setting. The proof of Lemma~\ref{bbb1} will be postponed to
the end of this subsection.

\begin{lem}\label{bbb1}
{\rm
There exists  $C_6 > 0$ so that, for $\varepsilon$ small enough,
$$t_\varepsilon(w) \geq C_6^{-1}\, \varepsilon^{-1} \int_{\mathbb R} |w|^2
ds, \qquad \forall w\in {\mathcal H}_0^1(\mathbb R).$$
}
\end{lem}

The proof of the next theorem is similar to the proof of
Theorem~\ref{aaa4}; it is enough to take into account   Lemma~\ref{eee4}, and then Lemma~\ref{bbb1}
instead of Lemma~\ref{ddd10}. Recall that ${\mathcal L}$ is the subspace  generated by functions
$w(s) u_0(y)$ with $w \in \LL^2(\R)$

\begin{teo}{\rm
Let $I=\mathbb R$. Then,
there exists $C_{10} > 0$ so that, for $\varepsilon$ small enough,
$$\left|\left| \left(- \Delta_{\varepsilon,c} -
\frac{\lambda_0}{\varepsilon^2 M^2} \Id \right)^{-1}
- \left(T_{\varepsilon,c}^{-1} \oplus 0\right) \right| \right| \leq C_{10}\,
\varepsilon^{3/2},$$
where  $0$ denotes the null operator on the subspace ${\mathcal L}^\perp$.
}
\end{teo}

As in the previous section, consider the self-adjoint operators
$$\hat{T}_{\varepsilon,c}:= \varepsilon J_\varepsilon T_{\varepsilon,c}
J_\varepsilon^{-1},$$
where $J_\varepsilon: \LL^2(\mathbb R) \to \LL^2(\mathbb R)$ is the previously discussed
unitary operator generated by the dilation
$s\mapsto s \varepsilon^{1/2}$.

\begin{teo}\label{ddd9}
{\rm
For $\varepsilon \to 0$ one has
$$\left| \left| \hat{T}_{\varepsilon,c}^{-1} - T^{-1} \right| \right| \to 0,
$$
where $T$ is the operator~\eqref{eqweo}.
}\end{teo}

As in the Section~\ref{qqq3},
we have that $\varepsilon J_\varepsilon W_\varepsilon(s)  J_\varepsilon^{-1}$ equals
$$
 \lambda_0 \,\zeta_\varepsilon(\varepsilon^{1/2} s, y) \left[ M^{-3} s^2+  \rho
(\varepsilon^{1/2} s) s^3
\varepsilon^{1/2} \right] + \varepsilon\, \vartheta(\varepsilon^{1/2} s) +
\varepsilon c.$$
Again, since
$\zeta_\varepsilon(\varepsilon^{1/2} s, y) \to 1$ uniformly as 
$\varepsilon \to 0$,
the proof of  Theorem~\ref{ddd9}
is similar to the proof of Theorem~$1.3$ in \cite{Frie}, and  details will
be skipped.

\

\noindent{\bf Proof of Lemma~\ref{bbb1}:}
Theorem~\ref{ddd9} guarantees that
$$\varepsilon^{-1} \left| \left| T_{\varepsilon,c}^{-1} \right| \right| \to
\left| \left| T^{-1} \right| \right|
\qquad (\varepsilon \to 0),$$
and so there exists $C_6> 0$ so that
$$\left| \left| T_{\varepsilon,c}^{-1} \right| \right| \leq C_6
\,\varepsilon.$$ The proof is complete.
$\cq$

\appendix
\numberwithin{equation}{section}

\section{Proof of Theorem~\ref{teoap}}

It will be shown that, for
$\varepsilon$ small enough, there exists $C_5>0$ so that
$$\left\|\hat{G}_\varepsilon^{-1}
- G_\varepsilon^{-1}\right \| \leq C_5\, \varepsilon.$$

Remember that $c > \|v\|_\infty + (1 / M^2)\|
k(s)^2 / 4\|_\infty$, thus, there exists a number $d > 0$ so that
$c = \|v\|_\infty + (1 / M^2)\|
k(s)^2 / 4\|_\infty + d$.

Since $\zeta_\varepsilon \to 1$ uniformly as $\varepsilon \to 0$,  there
exist  $\varepsilon_1 > 0$ and numbers
$\sigma_1, \sigma_2 > 0$ so that  $\sigma_1 \leq \beta_\varepsilon \leq
\sigma_2$,
for all $ \varepsilon < \varepsilon_1$. Thus,
$$\hat{g}_\varepsilon(\psi) \geq \sigma_1 d \| \psi\|^2 \qquad
\hbox{and} \qquad g_\varepsilon(\psi) \geq  d \| \psi\|^2,$$
for all $\varepsilon < \varepsilon_1$.
Consequently,
$$\|\hat{G}_\varepsilon^{-1}\| \leq \frac{1}{\sigma_1 d}
\qquad
\hbox{and} \qquad \|G_\varepsilon^{-1}\| \leq \frac{1}{d},$$
for all $\varepsilon < \varepsilon_1$.

Since $k, h \in \LL^\infty(\mathbb R)$, $y \in S$ and  $S$ is a bounded
region,
there exist $\varepsilon_0 > 0$ ($\varepsilon_0 < \varepsilon_1$)
and~$C_1, C_2 >0$ so that
$$\left|\left( \frac{1}{\zeta_\varepsilon} - 1  \right)\right| =
\left|\frac{\varepsilon\, k(s)h(s) (y \cdot 
z_\alpha(s))}{\zeta_\varepsilon}\right| \leq
C_1\, \varepsilon,$$
and
$$c\,|(\zeta_\varepsilon - 1)| \leq C_2\, \varepsilon,$$
for all $\varepsilon < \varepsilon_0$.
Under such conditions we have
\begin{eqnarray*}
& &
|\hat{g}_\varepsilon(\psi) - g_\varepsilon(\psi)| \\
& \leq  &
\int_{I \times S}
\left|\left(
\frac{1}{\zeta_\varepsilon} - 1 \right)\right| \left|\psi' - \psi
\,\frac{h'}{h}+ (\nabla_y \psi \cdot R y ) (\tau+\alpha') -
(\nabla_y \psi \cdot  y ) \frac{h'}{h} \right|^2
dy ds \\
& + &
\int_{I \times S} c |(\zeta_\varepsilon-1)| |\psi|^2\,ds dy \\
&\leq &
C_1  \varepsilon \int_{I \times S}
\left|\psi' - \psi \,\frac{h'}{h}+ (\nabla_y \psi \cdot R y ) (\tau+\alpha') -
(\nabla_y \psi \cdot  y ) \frac{h'}{h} \right|^2
+ C_2 \varepsilon \int_{I \times S} |\psi|^2 dy ds \\
& \leq &
C_3 \varepsilon g_\varepsilon(\psi)
\end{eqnarray*}
for some  $C_3> 0$. Hence,
$$(1 - C_3 \varepsilon) g_\varepsilon(\psi) \leq
\hat{g}_\varepsilon(\psi) \leq  (1 + C_3 \varepsilon)
g_\varepsilon(\psi),$$
for all $\varepsilon < \varepsilon_0$.
The first inequality implies that it is possible to find $\varepsilon_0' >
0$ ($\varepsilon_0' < \varepsilon_0$)
and a constant  $C_4>0$ so that
$$g_\varepsilon (\psi) \leq  C_4\, \hat{g}_\varepsilon(\psi),$$
for all $\varepsilon < \varepsilon_0'$.

By Schwarz's Inequality for bilinear forms, we have
$$|\hat{g}_\varepsilon(\psi_1, \psi_2)| \leq
[\hat{g}_\varepsilon(\psi_1)]^{1/2}\,[\hat{g}_\varepsilon(\psi_2)]^{1/2},$$
$$|g_\varepsilon(\psi_1, \psi_2)| \leq
[g_\varepsilon(\psi_1)]^{1/2}\,[g_\varepsilon(\psi_2)]^{1/2},$$
for all $\psi_1, \psi_2 \in {\mathcal H}_0^1(\mathbb R \times S)$.
Thus, by using the above estimates, for each pair $\psi_1, \psi_2 \in
{\mathcal H}_0^1(\mathbb R \times S)$
we have
\begin{eqnarray*}
\big|\big\langle  \hat{G_\varepsilon}^{1/2} \psi_1,
\hat{G_\varepsilon}^{1/2} \psi_2   \big\rangle
&-&
\big\langle  G_\varepsilon^{1/2} \psi_1, G_\varepsilon^{1/2}
\psi_2  \big \rangle\big| \\
& = &
\big|\hat{g}_\varepsilon  (\psi_1, \psi_2) - g_\varepsilon 
(\psi_1, \psi_2)\big| \\
& \leq &
C_3 \,\varepsilon\, [g_\varepsilon(\psi_1)]^{1/2}\,
[g_\varepsilon(\psi_2)]^{1/2}\\
& \leq &
C_3 \sqrt{C_4}\, \varepsilon\, [g_\varepsilon(\psi_1)]^{1/2}\,
[\hat{g}_\varepsilon(\psi_2)]^{1/2}.
\end{eqnarray*}

By picking $\psi_1 = G_\varepsilon^{-1} f$, $\psi_2=
\hat{G}_\varepsilon^{-1} g$, where
$f,g \in \LL^2(\mathbb R \times S)$ are arbitrary, we obtain
\begin{eqnarray*}
\Big|\big\langle \hat{G_\varepsilon}^{-1}f, g \big\rangle &-&
\big\langle G_\varepsilon^{-1}f, g \big\rangle \Big| \\
& \leq &
C_3 \sqrt{C_4}\, \varepsilon \left[\big \langle \hat{G}_\varepsilon^{-1} g, g\big \rangle
\,\big\langle G_\varepsilon^{-1} g, g \big\rangle  \right]^{1/2} \\
& \leq &
\frac{C_3 \sqrt{C_4}}{d \sqrt{\sigma_1}} \,  \varepsilon\, \|f\| \: \|g\|,
\end{eqnarray*}
for all $\varepsilon < \varepsilon_0'$.
Therefore,
$$\left\|\hat{G}_\varepsilon^{-1}
- G_\varepsilon^{-1} \right\| \leq C_5\, \varepsilon,$$
for all $\varepsilon < \varepsilon_0'$, with $  C_5 ={C_3
\sqrt{C_4}}/({d \sqrt{\sigma_1}})$. This completes the proof of the
theorem.

\

\subsection*{Acknowledgments} CRdeO thanks the partial support by CNPq (Brazil). AAV thanks the financial support by PNPD-CAPES (Brazil).

\end{document}